%% file: main.tex
\documentclass[conference]{IEEEtran}
\IEEEoverridecommandlockouts
\usepackage{cite}
\usepackage{amsmath,amssymb,amsfonts}
\usepackage{algorithmic}
\usepackage{graphicx}
\usepackage{textcomp}
\usepackage{xcolor}
\usepackage{cite}
\usepackage{amsmath,amssymb,amsfonts}
\usepackage{algorithmic}
\usepackage{graphicx}
\usepackage{caption}
\usepackage{textcomp}
\usepackage{xcolor}
\usepackage{svg}
\usepackage{afterpage}
\usepackage{booktabs}
\usepackage{float}
\usepackage{subfigure}
\usepackage{enumerate}
\usepackage{multirow}
\usepackage{todonotes}
\usepackage{soul}
\renewcommand{\baselinestretch}{0.985}
\usepackage[citecolor=blue,colorlinks=true,linkcolor=blue]{hyperref}%
\def\BibTeX{{\rm B\kern-.05em{\sc i\kern-.025em b}\kern-.08em
    T\kern-.1667em\lower.7ex\hbox{E}\kern-.125emX}}
\begin{document}

\title{Real-time zero-day Intrusion Detection System for Automotive Controller Area Network on FPGAs}

\author{\IEEEauthorblockN{Shashwat Khandelwal \& Shanker Shreejith}
\IEEEauthorblockA{ Reconfigurable Computing Systems Lab, Electronic \& Electrical Engineering\\
Trinity College Dublin, Ireland\\
Email: \{khandels, shankers\}@tcd.ie}}


\maketitle

\input{abstract.tex}

\input{introduction.tex}
\input{background.tex}
\input{architecture.tex}
\input{results.tex}

\input{conclusion.tex}
\bibliography{references}
\bibliographystyle{ieeetr}

\end{document}

%% file: abstract.tex
\begin{abstract}
Increasing automation in vehicles enabled by increased connectivity to the outside world has exposed vulnerabilities in previously siloed automotive networks like controller area networks (CAN).
Attributes of CAN such as broadcast-based communication among electronic control units (ECUs) that lowered deployment costs are now being exploited to carry out active injection attacks like denial of service (DoS), fuzzing, and spoofing attacks.
Research literature has proposed multiple supervised machine learning models deployed as Intrusion detection systems (IDSs) to detect such malicious activity; however, these are largely limited to identifying previously known attack vectors.
With the ever-increasing complexity of active injection attacks, detecting zero-day (novel) attacks in these networks in real-time (to prevent propagation) becomes a problem of particular interest.
This paper presents an unsupervised-learning-based convolutional autoencoder architecture for detecting zero-day attacks, which is trained only on benign (attack-free) CAN messages.
We quantise the model using Vitis-AI tools from AMD/Xilinx targeting a resource-constrained Zynq Ultrascale platform as our IDS-ECU system for integration. 
The proposed model successfully achieves equal or higher classification accuracy ($>$ 99.5\%) on unseen DoS, fuzzing, and spoofing attacks from a publicly available attack dataset when compared to the state-of-the-art unsupervised learning-based IDSs.
Additionally, by cleverly overlapping IDS operation on a window of CAN messages with the reception, the model is able to meet line-rate detection (0.43\,ms per window) of high-speed CAN, which when coupled with the low energy consumption per inference, makes this architecture ideally suited for detecting zero-day attacks on critical CAN networks. 
\end{abstract}

%% file: introduction.tex
\section{Introduction}\label{sec:introduction}
Automotive electronic systems have increasingly become complex to support novel capabilities for safety, infotainment, and comfort. 
Over a hundred electronic computing units (ECUs), sensors, and actuators are common in modern cars which are interconnected through multiple network protocols to exchange sensor information, control sequences, and actuation commands. 
Among the network protocols, Controller Area Networks (CAN)~\cite{CanBosch} continues to be the most widely used network protocol for in-vehicle networks owing to its cost-effective nature and ease of use in control applications.
While early ECUs and software functions were developed as siloed functions with limited connectivity to the external world, many novel capabilities rely on connectivity to infrastructure and other vehicles to enable remote monitoring and control of specific capabilities for diagnostics, over-the-air upgrades, and features. 
However, many recent researchers have shown that enabling external connectivity opens up new avenues for injecting malicious code/messages into previously siloed networks~\cite{nie2017free,iehira2018spoofing,cai20190}.
Such attacks are largely enabled by the lack of inherent security and authentication mechanisms in CAN and similar automotive network protocols~\cite{miller2015remote,enev2016automobile}. 

\begin{figure}[t!]
  \centering
  \includegraphics[scale = 0.4]{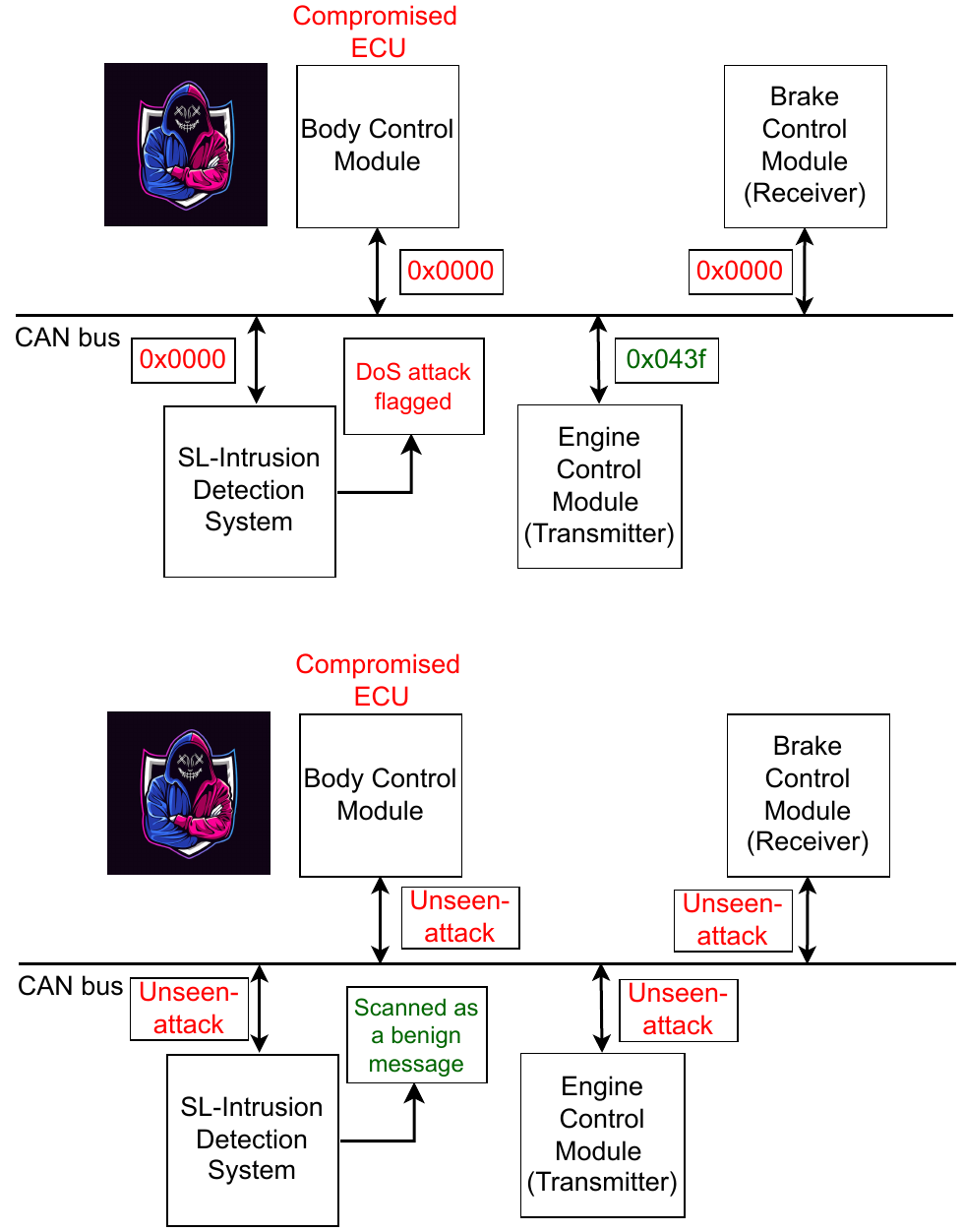} 
  \caption{An illustration of an zero-day attack  launched through a compromised ECU left undetected by an IDS. Top figure shows a supervised learning (SL) based IDS flagging the known DoS attack on CAN bus, while in bottom figure, the IDS is unable to flag the attack on the CAN bus as the compromised ECU uses zero-day attack techniques.}
  \label{fig:attack}
\end{figure}

Multiple intrusion detection approaches have been proposed in the literature that aim to detect such threats on the network and for allowing critical systems to enter into a `safe working' mode when such threats are detected. 
Early intrusion detection systems (IDS) proposed were rule/specification based, which utilised a set of rules to compare known attack signatures to patterns captured from current network parameters/messages to detect unusual activity~\cite{larson2008approach,miller2013adventures,studnia2018language}. 
Recently, supervised and unsupervised machine learning (ML) approaches have shown significant improvement in the detection accuracy of such threats and the ability to adapt to newer attack vectors~\cite{narayanan2015using, alshammari2018classification, yang2019tree, song2020vehicle, tariq2020cantransfer} without incurring overheads of rule-based approaches. 
Although utilising supervised learning models as IDSs have been more successful in detecting known attacks on which they are trained, they lack the capability of detecting new attack vectors (or zero-day attacks)~\cite{9974508}. 
IDS approaches that rely on unsupervised learning models have been shown to be successful in detecting zero-day (novel) attacks while simultaneously classifying known attacks successfully~\cite{yang2021mth,seo2018gids}. 
With increasing connectivity and automation in vehicles, the potential for novel attack vectors is also likely to increase, making the detection of zero-day attacks, as well as known vectors a priority~\cite{upstreamreport}.
Despite the promising performance of unsupervised learning models, integrating them within a vehicular network is non-trivial; firstly, most modern vehicles use multi-standard network architecture with different performance, capabilities, and medium access schemes requiring complex deployment strategies.
Secondly, most (supervised and unsupervised) approaches in the literature rely on ECUs with loosely-coupled dedicated accelerators (eg. GPUs) for near-line-rate detection of threats and/or dedicated IDS ECUs which could offset energy budget, weight/cabling constraints, and other overheads.
Hybrid FPGA-based ECUs have shown to be a promising platform to achieve consolidation of multiple functions with clear isolation between them on the same die. 
Additionally, specialised accelerators could be closely coupled to software functions to improve their throughput and energy efficiency~\cite{vipin2014mapping}.

In this paper, we propose a convolutional autoencoder model (CAE) as an IDS function, integrated through a hardware-efficient ML engine to accelerate the IDS operation while offering complete isolation to the ECU software function. 
The CAE model is custom-quantised to right-size the model using AMD/Xilinx's Vitis-AI framework~\cite{xilinxvitis} and is deployed as an Advanced eXtensible Interface (AXI) slave peripheral of the ECU for enabling IDS capabilities at the ECU-level. 
This integration allows the software tasks on the ECU to invoke and fully control the operation of the IDS accelerator through APIs, similar to compute offloads enabled by AUTOSAR abstractions. 
The key contributions of this paper are as follows: 
\begin{enumerate}[1.]
    \item An unsupervised learning model (quantised convolutional autoencoder) based-IDS for automotive CAN achieving state-of-the-art classification accuracy across multiple unseen attack vectors demonstrating the capacity to detect zero-day attacks.
    \item A tightly-coupled ECU architecture that integrates a custom ML-based accelerator for offloading IDS tasks in full isolation.
    \item Quantify the performance and energy savings of the proposed unsupervised model and its integration using the open CAN dataset. Our results show that the proposed IDS achieves significant improvements in terms of per-message processing latency and power consumption against the state-of-the-art unsupervised learning-based IDSs proposed in the research literature.
\end{enumerate}

We evaluate our approach using the openly available CAR Hacking dataset~\cite{song2020vehicle} for training and validation of the model on normal CAN messages and testing it across multiple unseen attack vectors captured from an actual vehicle with a block of CAN IDs used as an input feature to improve the detection performance. 
Our experiments show that the proposed CAE-based IDS (referred to as CAE-IDS) achieves an average accuracy of 99.61\% across multiple unseen attack vectors such as Denial of Service (DoS), Fuzzy, and spoofing (RPM and Gear) attacks, identical to or exceeding the detection accuracy achieved by state-of-the-art \text{GPU- and CPU-based} implementations. 
The tightly integrated ECU architecture processes a block of 100 CAN messages in 0.43ms to enable real-time detection and reduces the power consumed by $\approx$ 2$\times$ compared to state-of-the-art IDSs proposed in the research literature. 

The remainder of the paper is organised as follows. Section~\ref{sec:background} discusses the background information and state-of-the-art research in this area; section~\ref{sec:architecture} describes the proposed CAE model and design choices for the implementation; section~\ref{sec:experiments} outlines the experiment setup and results; and we conclude the paper in section~\ref{sec:conclusion}.


%% file: background.tex
\section{Background and Related works}\label{sec:background}
\subsection{Controller Area Network}
\begin{figure}[t!]
  \centering
  \includegraphics[width=\linewidth]{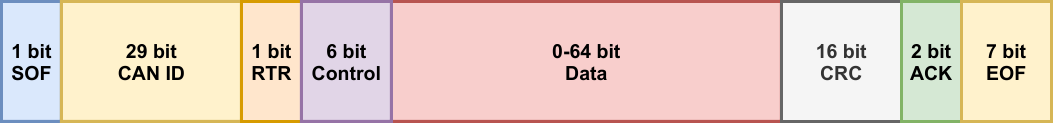}
  \caption{Frame format of an extended frame CAN message. In the case of a standard CAN message, the ID field contains the 11-bit identifier.}
  \label{fig:1}
\end{figure}

In-vehicle networks enable distributed ECUs to exchange control and data messages to achieve the global functions of the vehicle. 
Multiple protocols are used in vehicular systems to cater to different functions based on their criticality and to optimise the cost of E/E systems. 
CAN~\cite{CanBosch} and CAN-FD~\cite{hartwich2012can} continue to be the most widely used protocol today due to their lower cost, flexibility, and robustness. 
Figure~\ref{fig:1} illustrates the bit-field definition of a CAN data frame, with each segment providing some function in the network operation. 
The CAN ID (11-bit base, 29-bit extended format) is a unique identifier assigned to each message and by extension, defines its priority for transmission on the shared bus.
The message itself is contained in the data field.

The broadcast CAN bus uses a bit-wise arbitration method to control medium access to the bus using the CAN ID allocated to each message. 
CAN also support multiple data rates (125\,Kbps to 1\,Mbps) and multiple modes of operation (1-wire, 2-wire) to cater to a range of critical and non-critical functions in vehicles. 
Despite this robustness, CAN is inherently insecure: there is no built-in mechanism in the network to authenticate the transmitter, receiver, or the message content itself~\cite{8658720}.  
This makes CAN vulnerable to simple and efficient attacks like message sniffing, fuzzing, spoofing, replay attacks, and Denial of Service (DoS) attacks~\cite{mukherjee2016practical,enev2016automobile,koscher2010experimental,palanca2017stealth}. 

\subsection{IDSs for CAN}
Researchers have explored different flavours of IDSs from rule/flow-based approaches to machine learning-based methods to address CAN's vulnerabilities.
Rule-based approaches are generally classified into flow-based and payload-based. 
Flow-based approaches identify traits like message frequency and/or interval for the network to detect abnormalities~\cite{vuong2015performance}, payload-based approaches use the data segment in CAN frames to detect abnormal sequences of instructions and or data~\cite{stabili2017detecting}.
Hybrid schemes use both the message timing/frequency and the payload information to capture a more holistic view of the network, allowing them to extract specific signatures of transmitting ECUs, receiving ECUs, and messages~\cite{weber2018embedded}. 
For instance, the fingerprint-based approach uses low-level electrical signal levels and timing of signals with the message contents to identify potential intrusions when large deviations are observed~\cite{cho2016fingerprinting}.
Machine learning-based IDSs utilise classification approaches, sequential techniques, and deep learning-based schemes to achieve better generalisation and threat detection in CAN.

\begin{table}[t]
\centering
\caption{Input features used by the IDSs \& IPSs proposed in the research literature.}
\begin{tabular}{llc}
\toprule
\textbf{Models} & \textbf{Input Features} & \textbf{Learning}                 \\
\midrule
\textbf{GIDS}~\cite{seo2018gids}            & CAN ID   & UL                                       \\
DCNN~\cite{song2020vehicle}            & CAN ID   & SL                                       \\
Rec-CNN~\cite{desta2022rec}         & CAN ID   & SL                  \\
\textbf{iForest}~\cite{de2021efficient}         & Data Field  &  UL                                 \\
MLIDS~\cite{desta2020mlids}           & CAN ID + Data Field  & SL                         \\
\textbf{NovelADS}~\cite{agrawal2022novelads}        & CAN ID + Data Field  & UL                        \\
TCAN-IDS~\cite{cheng2022tcan}        & CAN ID + Data Field & SL                         \\
\textbf{MTH-IDS}~\cite{yang2021mth}         & CAN ID + Data Field  & UL \\
HyDL-IDS~\cite{lo2022hybrid}           & CAN ID + Data Field + DLC  & SL                       \\
GRU~\cite{ma2022gru}               &  CAN ID + Data Field + DLC & SL \\ 
\textbf{QCAE-IDS (proposed)}         & CAN ID & UL \\
\bottomrule
\end{tabular}
\label{table:inpfeatures}
\end{table}

\subsection{Machine Learning-based IDSs}
Most ML-based IDSs fall under the hybrid category since they use a combination of input features, as shown in Table~\ref{table:inpfeatures}.
Here, supervised and unsupervised learning are denoted as \emph{SL} \& \emph{UL} respectively.
Multiple supervised learning-based IDSs have been proposed and shown to be highly successful across all metrics~\cite{song2020vehicle,9974508,ma2022gru,desta2020mlids}. 
Unsupervised learning models on the other hand have also been shown to generalise fairly well on benign datasets to detect unseen attacks including zero-day. To determine their efficiency, they are tested against an attack dataset to detect known attacks like DoS, fuzzing, and spoofing~\cite{agrawal2022novelads,yang2021mth,de2021efficient,seo2018gids}.    
In~\cite{seo2018gids}, the authors propose a GAN-based unsupervised learning IDS and achieve an average accuracy of 97.5\% for the unseen (DoS, fuzzing, and spoofing) attacks. 
In~\cite{de2021efficient}, the authors use an isolation forest unsupervised anomaly detection algorithm as an intrusion prevention system (IPS) to detect fuzzing and spoofing (RPM \& Gear) attacks and mark the message as an error preventing its propagation to other ECUs; however, this can cause multiple messages to be dropped from the bus in case of false positives or DoS attacks.
In~\cite{yang2021mth}, the authors proposed a stacked anomaly-based IDS for detecting zero-day attacks and report a slightly lower average F1 score of 96.3\% on the attacks included in the Car hacking dataset.
In~\cite{agrawal2022novelads}, the authors propose an unsupervised learning-based CNN+LSTM autoencoder architecture for incoming message reconstruction and classifying them as benign and attacks using thresholding techniques.
They report high accuracy of 99.9\% for all attacks but their high message processing latency makes them unsuitable for real-time detection. 

The key challenge of ML-based approaches is their deployment as an in-vehicle ECU. 
Most approaches rely on high-performance GPUs to meet the inference deadlines~\cite{seo2018gids,song2020vehicle,desta2020mlids,ma2022gru,agrawal2022novelads,cheng2022tcan}, while others rely on dedicating full ECUs for IDS~\cite{yang2021mth,de2021efficient}; both approaches incur additional overheads in energy budget, integration, and weight, making them less suited for distributing IDS among different network segments. 

\subsection{Autoencoders as anomaly detectors}
Autoencoders are neural networks used for the faithful reconstruction of the input features they are trained on following an \emph{encoder-decoder} architecture.
During the training process, the encoder block extracts low-level information from high-level input feature representation, storing the most vital information in its \emph{latent} space (also called \emph{bottleneck}).
The encoder network is represented as a standard neural network function:
\begin{align*}
    z = \sigma (Wx + b)
\end{align*}
The activation function, input, weights, and biases of the encoder are represented by $\sigma$, \emph{x}, \emph{W}, and \emph{b} respectively, with \emph{z} representing the learned latent space.
A smaller latent space prevents the model from overfitting on the training data.
The decoder block then uses this encoded information from the latent space to correctly reconstruct the provided input.
The decoder network can also be represented as a neural network similarly:
\begin{align*}
    \Tilde{x} = \Tilde{\sigma} (\Tilde{W}z + \Tilde{b})
\end{align*}
The activation function, input, weights, and biases of the decoder are represented by $\Tilde{\sigma}$, \emph{z}, \emph{$\Tilde{W}$} and \emph{$\Tilde{b}$} respectively, with \emph{$\Tilde{x}$} representing the reconstructed output.
The training stops when the autoencoder starts predicting the input with an acceptable level of loss.
Autoencoders are used for a variety of applications like dimensionality reduction~\cite{sakurada2014anomaly}, image denoising~\cite{gondara2016medical}, generation of data~\cite{bengio2013generalized} and anomaly detection~\cite{govorkova2022autoencoders}.

In the case of anomaly detection, autoencoders work by reconstructing an output significantly different from the input, leading to a higher loss for the unseen, unpredictable or out-of-sequence data input.
Using thresholding techniques, an optimum threshold of the loss is set using the normal data.
Any data point during testing which gives a higher loss than the threshold is flagged as an anomalous data point.

%% file: architecture.tex
\section{System architecture}\label{sec:architecture}
\subsection{CAE model as unsupervised IDS}
To determine the best-unsupervised ML model for IDS, we profiled different ML architectures like dense autoencoders, convolutional autoencoders, and a combination of both to find a model with high classification accuracy and low computational complexity.
Among the candidate architectures, we observed that CAE's were effective in detecting attacks when trained using a window of CAN-IDs as the input feature. 

\begin{figure}[t!]
    \centering
    \includegraphics[scale = 0.48]{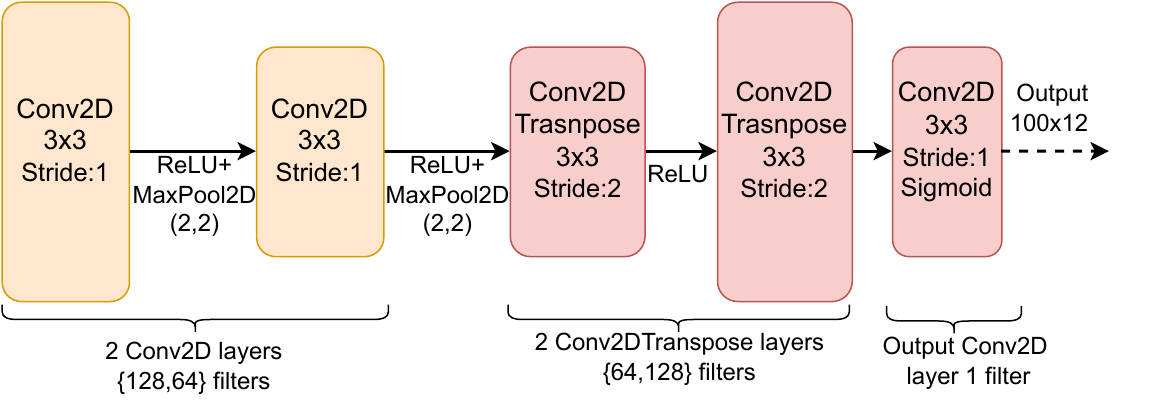}
    \caption{The proposed convolutional autoencoder model as a zero-day attack detection IDS.}
    \label{fig:model}
\end{figure}

The autoencoder is comprised of an encoder-decoder architecture with 2 \emph{Conv2D} layers and \emph{relu} activation in the encoder with max-pooling layers in between to reduce features.
The decoder comprises 2 \emph{Conv2DTranspose} (or Deconvolution) layers with \emph{relu} activation to reconstruct the input from the learned latent space during the training process.
The final layer of the decoder is comprised of a \emph{Conv2D} layer with a single filter and \emph{sigmoid} activation to output the final binarized reconstructed message as shown in Figure~\ref{fig:model}. 
The model is defined in TensorFlow (TF) using standard TF functions and node definitions.

As the input feature, a sequence of \textit{n = 100} CAN IDs (in binary) was chosen, capturing the sequence of messages exchanged on a normal CAN network. 
At 100\% bus utilisation, this corresponds to a time window of 1\,ms (at the peak rate of 10000 messages per second) on a high-speed CAN network. 
With a lower number of messages per block (e.g., 12, 50 from our exploration), we observe that the model has fewer input data to learn the intrinsic features from, reducing the efficiency of the classifier. 
Similarly, increasing the window (or block size $>$ 100) resulted in much higher worst-case detection latency for the attack, due to the time required to accumulate the required number of messages.



\subsection{Design space exploration for CAE parameters}
\label{subsec:DSE}

\begin{figure}[t!]
    \centering
    \includegraphics[scale = 0.35]{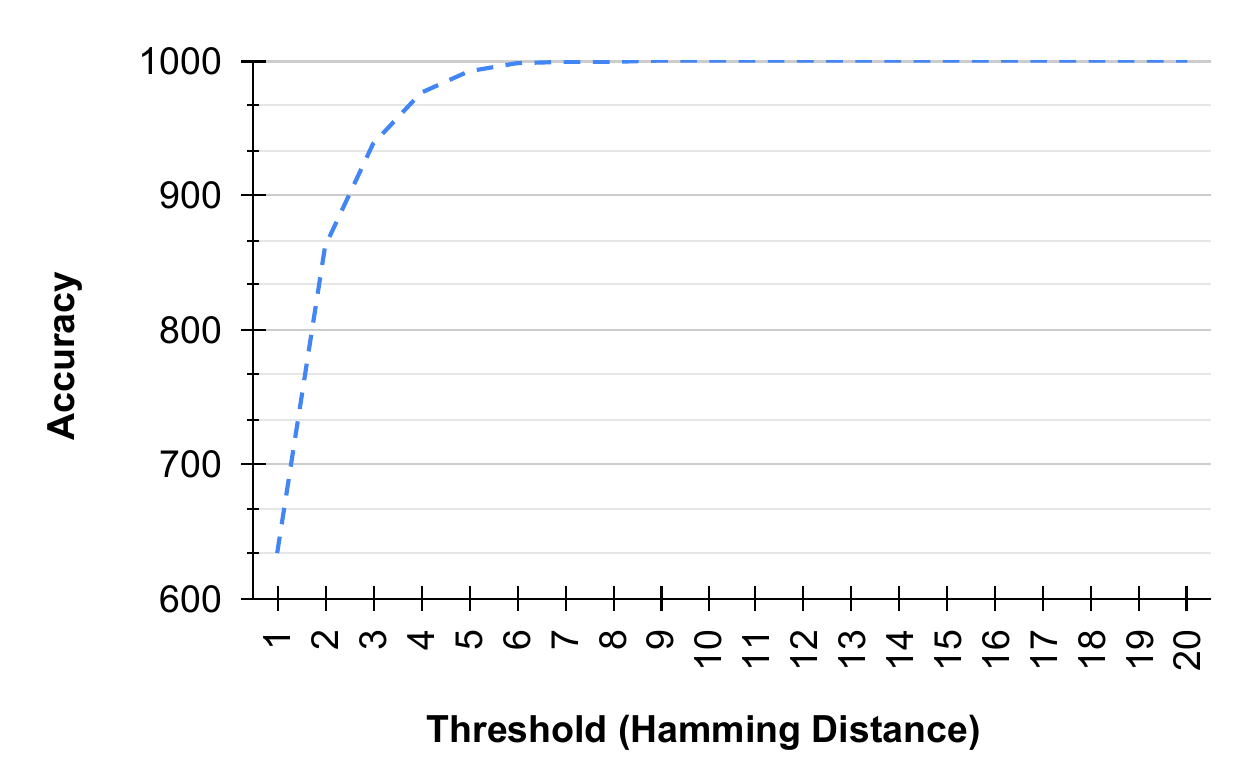}
    \caption{Hamming distances of the reconstructed messages on the benign (attack-free) dataset. 1000 blocks of CAN messages were used in testing for this dataset.}
    \label{fig:normal_dist}
\end{figure}

\begin{figure}[t!]
    \centering
    \includegraphics[scale = 0.35]{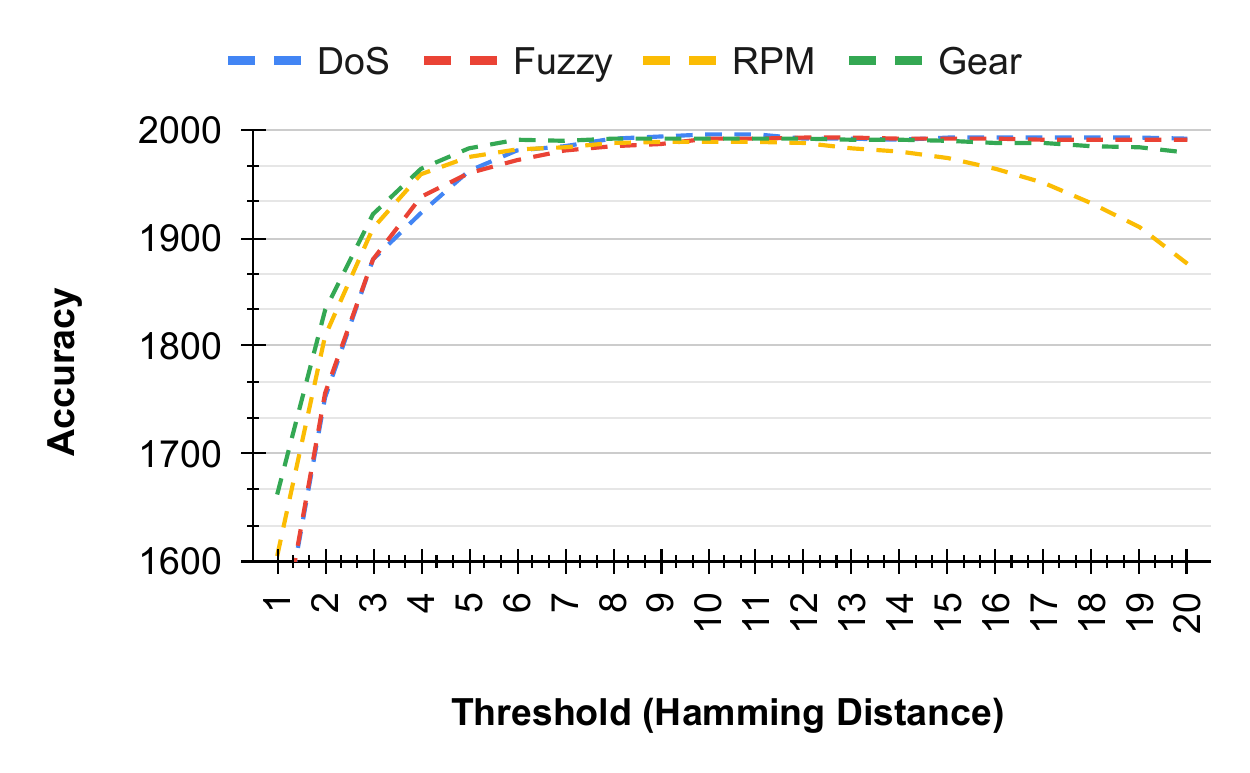}
    \caption{Hamming distances of the reconstructed messages on the DoS, fuzzing and spoofing attack datasets. 2000 blocks of CAN messages were used in testing for each attack dataset.}
    \label{fig:attack_dist}
\end{figure}

To determine the optimal configuration of our CAE, we explored different design configurations with different layers and filter sizing, iterating through multiple training and validation steps.
For the \emph{Conv2D} and \emph{Conv2DTranspose} layers we experimented with different filter sizes \{128,\,64,\,32\} in the encoder and decoder and found that the configuration \{128,\,64\} \& \{64,\,128\} gave the highest classification performance while keeping the computational complexity of the model low (among the other configurations).
To classify the reconstructed messages as benign or attack we chose the \emph{hamming distance} metric to compute the distance (threshold) between the original and predicted binary vectors.
When the network trained on normal messages encounters attack messages in the input stream, the reconstructed message results in a much larger hamming distance (greater than the classification threshold) as the model has not seen these messages during training.
To arrive at the optimal classification threshold, we evaluate the accuracy of the IDS on the benign dataset by varying the threshold from 0 to 20. 
We find that for threshold values from 0 to 9, the number of false positives decreases gradually, while for values of \emph{10} and above, we get zero false positives for classifying normal messages, leading to \emph{10} as the optimal classification threshold. 
Subsequent validation on the attack dataset showed that the chosen threshold achieves very good classification results across all four attack classes, detailed results of which are reported in section~\ref{sec:experiments}.
Figures~\ref{fig:normal_dist} \&~\ref{fig:attack_dist} show the accuracy of the model for all the thresholds on the benign and the attack datasets during testing.

With the parameter configurations and architecture for the proposed convolutional autoencoder architecture determined through the design space exploration, the weights, biases, and activations are 8-bit quantised using AMD/Xilinx's Vitis-AI framework~\cite{xilinxvitis} to generate the quantised version of our CAE model (referred to from hereon as QCAE). 
Subsequently, we explored different hardware accelerator models using different configurations of Vitis-AI's deep learning processing unit (DPU) IP core~\cite{xilinxdpu} to deploy the QCAE IDS.
Based on the resources available on the target Ultrascale+ device (XCZU7EV), three different DPU configurations (B512, B1152, B4096) were evaluated to compare the latency and energy consumption of the model, with B4096 DPU providing the best tradeoff (results in  section~\ref{sec:experiments}).
For our inference results and comparisons, the B4096 DPU was chosen as the accelerator.

\subsection{Integration to the ECU-IDS}

\begin{figure}[t!]
    \centering
    \includegraphics[scale = 0.55]{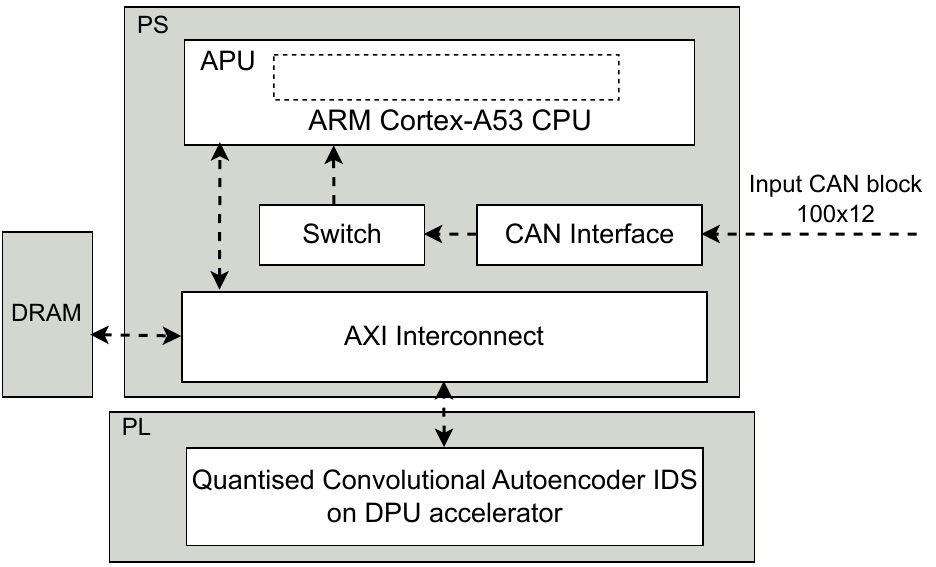}
    \caption{Proposed system architecture of the integrated IDS. The quantised CAE model is accelerated on the PL part of the FPGA device.}
    \label{fig:datapath}
\end{figure}

Figure~\ref{fig:datapath} shows the proposed ECU architecture of the IDS-enabled ECU on a Xilinx Zynq XCZU7EV device. 
The multiple ARM processor cores in the PS section are connected to a host of hardened memory-mapped peripheral logic and interface protocols, allowing seamless integration of standard ECU functionality as a baremetal application or with an operating system.
Custom accelerators can be deployed in the programmable logic (PL) region, and integrated as a custom peripheral of the PS using high or low-performance Advanced eXtensible Interface (AXI) ports, making them accessible from the ECU application.
In our proposed ECU architecture, standard ECU function(s) are mapped as software tasks onto one or more of the ARM cores on the PS.
The operating system or baremetal application provides relevant drivers and APIs for accessing the PS peripherals and the PL accelerators, abstracting away low-level details of these blocks to create an AUTOSAR-compliant architecture~\cite{fons2012fpga}.
We propose to use the integrated CAN interface on the PS to handle the interfacing of ECU to the CAN bus, as shown in Figure~\ref{fig:datapath}.

The PL on the FPGA is programmed with Xilinx's DPU IP core, exported from the Vitis-AI framework. 
Vitis-AI generates the required wrappers for integrating the DPU core with the processor subsystem as well as the low-level drivers for enabling data exchange between the ARM cores and the DPU. 
The QCAE model was translated into an `xmodel' file that provides instructions to the DPU core; at runtime, the `xmodel' is loaded by the ARM cores during the ECU boot-up sequence which is subsequently invoked through a \emph{execute-async} command from the ECU task to execute the model whenever new CAN window is ready to be processed.
Ping-pong buffers are integrated to accumulate the block of CAN messages as they arrive, allowing a clean overlap between IDS execution and the subsequent block of message accumulation. 

\subsection{Dataset and Training}
\label{subsec:dataset}
We use the open Car Hacking dataset for training our model and to test its performance~\cite{canlink}. 
The dataset provides a labelled set of normal (benign) and attack messages which were captured via the Onboard Diagnostic (OBD) port in an actual vehicle, with attack messages injected in real-time. 
The dataset consisting of attack free set of messages is used to train the CAE helping it generalise on the normal flow of CAN traffic and sets of attack datasets with DoS, Fuzzing Spoofing message injections, allowing us to test the detection accuracy of the IDS against these \emph{previously unseen} attacks. 
We split the normal message dataset as 75:15:10 for training, validation, and testing respectively, allocating the large section to training and optimisation of the quantised network. From the four attack datasets used for testing, we extract 200,000 messages in sequence from each of them, resulting in 2000 blocks (block size = 100) of messages for testing. 
If the block comprises one or more attack messages the entire block is labelled as an attack block.
The performance of the model on the validation set during training ensures that it is not overfitting on the benign message dataset. 
The normal messages and attack messages are similarly pre-processed (as blocks of 100 messages) before training to mimic the dataflow the model will obtain as its input when integrated into the ECU. 
With 12-bit CAN IDs used in the dataset, the binarised input block has the shape \textit{\{100,12\}}, which is organised as the input FIFO/buffer when integrated with the PS.  
These blocks are then fed into the model as input for training and testing respectively.

The model was trained using the standard TensorFlow framework.
We used the \emph{adam} optimizer with \emph{mean squared error} (MSE) as the loss function. 
\begin{align*}
    Loss = \lvert \Tilde{X} - X \rvert ^2
\end{align*}
The learning rate was set to 0.001 to allow for slower learning which aids in reducing loss of accuracy when quantising the pre-trained model~\cite{wu2018training}. 
The model has 187,079 learning parameters and was trained for 100 epochs with a batch size of 64 on the normal (attack-free) dataset.
The best model in terms of validation loss was saved and exported as an \textit{'h5'} file.
This weights file is then fed into Vitis-AI for quantisation and post-quantisation optimisations (as discussed above) for final deployment on the DPU IP.

%% file: results.tex
\section{Experimental Results}\label{sec:experiments}
For training, we use the floating point variant of the unsupervised CAE model and use an Nvidia A6000 GPU for accelerating the training and validation, prior to quantisation through Vitis-AI.
To quantify the performance and energy benefits of tightly integrating CAE-IDS with ECU functions, we use a Zynq Ultrascale+ ZCU104 development board that features an XCZU7EV Ultrascale+ device with quad-core ARM A53 cores and dual-core ARM R5 cores on the PS as our target platform.
A standard Linux kernel with petalinux tools and VART interfaces enabled is used as the boot configuration for the ARM cores.
The VART APIs are used to program the model (`xmodel' binary) onto the DPUs (in PL) at startup and to trigger the execution of the model at runtime.
The A53 cores on the PS are configured to run at 1.2\,GHz peak. 
The DPUs use a 600\,MHz DSP core clock and are configured to run concurrently for all our tests at startup.
The Nvidia A6000 GPU has a base operating frequency of 1350\,MHz. 

For testing the IDS on FPGA with the `xmodel' binary loaded, we test the performance using the test split from all four attack datasets.  
We quantify the inference accuracy by evaluating the precision, recall, and F1 rates of our QCAE model.
We also quantify the per-block processing latency and average power consumption for performing IDS on each incoming CAN message on the ECU.
The results, in terms of inference accuracy and per-block processing latency, are compared against the state-of-the-art IDSs/IPSs proposed in the research literature.
We also compare the per-block processing latency and the per-inference energy consumption when the FP32 version of the proposed model is executed on an NVIDIA Jetson Nano 4GB GPU.
In the case of schemes where inference is performed on a block of CAN messages, we use these metrics along with the block size for the comparison.
We also compare our active power consumption against ML-IDS approaches in literature where power consumption has been reported.

\subsection{Accuracy}
To test the functional correctness of the model, we first compare the pre- \& post-quantisation inference performance of the model. 
We see an equal or higher performance for the DoS \& Gear spoofing attacks and a negligible drop in the F1 scores for the Fuzzing \& RPM spoofing attacks pre \& post quantisation as shown in table~\ref{table:qcae_perf}.
We first evaluate the performance of the model on the benign dataset to test for false positives.
Out of 1000 blocks of messages that are tested, we find all of them being classified as benign with zero false positives.
To demonstrate the ability to detect zero-day attacks, we test the classification performance of the model on unseen \textit{DoS, Fuzzy} and spoofing (\textit{Gear and RPM}) attacks from the Car Hacking dataset and compare them against the competing techniques (unsupervised learning techniques represented in bold in the comparison tables) from the literature. 
Table~\ref{table:confmatrix_mlp1} captures the classification performance of our model in isolation across our test set as a confusion matrix. 
As discussed in section~\ref{subsec:DSE}, the accuracies reported are obtained for the threshold value of \emph{10} for all datasets.
The misclassifications (false negatives) observed while testing on unseen attack datasets can be attributed to scenarios where the attack message flow closely resembles that of the normal messages within the observed block of messages at a given time. 

We compare the inference performance of our QCAE model integrated within the ECU against the state-of-the-art IDSs and IPSs proposed in the literature: GIDS~\cite{seo2018gids}, DCNN~\cite{song2020vehicle}, MLIDS~\cite{desta2020mlids}, HyDL-IDS~\cite{lo2022hybrid} NovelADS~\cite{agrawal2022novelads}, TCAN-IDS~\cite{cheng2022tcan}, IForest~\cite{de2021efficient}, MTH-IDS~\cite{yang2021mth}, GRU~\cite{ma2022gru} and Rec-CNN~\cite{desta2022rec} which are captured in tables~\ref{table:comp1} and ~\ref{table:comp2}, comparing them in terms of inference precision, recall, F1 score.
In~\cite{yang2021mth}, the authors report an average F1-score of 96.3\% for detecting zero-day attacks based on their stacked anomaly-based IDS.
In comparison, our QCAE achieves an average F1 score of 99.6\% for detecting the same set of unseen attacks.
When compared to other unsupervised learning-based techniques, for the \textit{DoS} attack QCAE performs better than ~\cite{seo2018gids} by 1.7\% in terms of the F1 score and only slightly less than~\cite{agrawal2022novelads} by 0.2\%.  
In the case of \textit{fuzzing} attack, the QCAE model performs better than~\cite{seo2018gids,de2021efficient} by 1.2\% \& 2\% respectively in terms of the F1 score and only slightly less than~\cite{agrawal2022novelads} by 0.5\%.
In the case of \textit{RPM spoofing} attack, our QCAE model performs better than~\cite{seo2018gids,de2021efficient} by 0.9\% \& 0.1\% respectively in terms of the F1 score and only slightly less than~\cite{agrawal2022novelads} by 0.4\%.
In the case of \textit{Gear spoofing} attack, our QCAE model performs better than~\cite{seo2018gids,de2021efficient} by 2.4\% \& 2.3\% respectively in terms of the F1 score and only slightly less than~\cite{agrawal2022novelads} by 0.3\%.
It can be observed that our QCAE model achieves comparable detection accuracy as state-of-the-art supervised learning models for all attacks as shown in  tables~\ref{table:comp1} and~\ref{table:comp2} respectively.

\begin{table}[t!]
\centering
\caption{Confusion matrix capturing the classification results of our QCAE on the attack-free (benign), DoS, Fuzzing and Spoofing (Gear, RPM) attacks. (Note the model is trained only on benign messages and does not see any attack message during the training phase.)}
    \scalebox{1}{
        \begin{tabular}{@{}llrr@{}}
        \toprule
            \textbf{Attack} & \textbf{Message Type}  & Predicted Normal  & Predicted Attack  \\ 
            \midrule
          \multirow{2}{*}{Benign} & True Normal & 1000 & 0\\ 
            & True Attack & NA & NA \\
          \multirow{2}{*}{DoS} & True Normal & 1082 & 2\\ 
            & True Attack & 2 & 914 \\ 
            \multirow{2}{*}{Fuzzy} & True Normal & 1190 & 3  \\
           &  True Attack & 5 & 802 \\ 
           
           \multirow{2}{*}{Gear} & True Normal & 823 & 0 \\
           &  True Attack & 11 & 1166 \\ 
           
           \multirow{2}{*}{RPM} & True Normal & 828 & 1 \\
           &  True Attack & 7 & 1164 \\ 
            \bottomrule
        \end{tabular}}
\label{table:confmatrix_mlp1}
\end{table}

\begin{table}[t!]
\centering
\caption{Inference accuracy metrics of CAE pre- and post-quantisation (QCAE) on all the attacks.} 
\scalebox{1}{
\begin{tabular}{@{}lllllll@{}}
\toprule
\textbf{Attack} & \textbf{Model}  & \textbf{Precision} & \textbf{Recall} & \textbf{F1} & \textbf{FPR} & \textbf{FNR} \\
\midrule
\multirow{2}{*}{DoS} & Pre-Q          & 99.56             & 99.89              & 99.73    & 0.37 & 0.11\\
& QCAE         &    99.78          &     99.78           &  99.78    & 0.18 & 0.22  \\
\multirow{2}{*}{Fuzzy} & Pre-Q          &   99.75           &    99.38       &   99.57  & 0.17 & 0.62  \\
& QCAE        &      99.63         &   99.38       &99.50& 0.25  & 0.62   \\ 
\multirow{2}{*}{RPM} & Pre-Q          &  100            &    99.15            &  99.57   & 0 & 0.85  \\
& QCAE        &       100       &       99.07         &   99.53  & 0  & 0.93  \\
\multirow{2}{*}{Gear} & Pre-Q          &       99.91       &     99.40      & 99.66    & 0.12 & 0.60   \\
& QCAE        &       99.91       &     99.40      & 99.66    & 0.12 & 0.60   \\ 
\bottomrule
\end{tabular}}
\label{table:qcae_perf}
\end{table}

\begin{table}[t!]
\centering
\caption{Accuracy metric comparison (\%) of our QCAE accelerator against the reported literature on the DoS and Fuzzing attacks.}
\scalebox{1}{
\begin{tabular}{@{}llllll@{}}
\toprule
\textbf{Attack}  & \textbf{Model} & \textbf{Precision} & \textbf{Recall} & \textbf{F1}  & \textbf{FPR} \\
\midrule
\multirow{5}{*}{DoS} & \textbf{GIDS}~\cite{seo2018gids}                  &   96.8              &   99.6        & 98.1  &   -   \\ 
& DCNN~\cite{song2020vehicle}                  & 100                & 99.89          & 99.95  & -     \\
& MLIDS~\cite{desta2020mlids}                  & 99.9                & 100          & 99.9  & -     \\
& HyDL-IDS~\cite{lo2022hybrid}                  & 100               & 100          & 100  & -     \\
& \textbf{NovelADS}~\cite{agrawal2022novelads}                  & 99.97               & 99.91          & 99.94  & -     \\
& TCAN-IDS~\cite{cheng2022tcan}                  & 100             & 99.97          & 99.98  & -     \\
& \textbf{iForest}~\cite{de2021efficient}                  & -                &   -       &  - &  -    \\ 
& GRU~\cite{ma2022gru}                  & 99.93             & 99.91               & 99.92  & -     \\
& \textbf{QCAE-IDS}                  & \textbf{99.78}             & \textbf{99.78}               & \textbf{99.78}  &  \textbf{0.18}    \\
\midrule
\multirow{5}{*}{Fuzzy} & \textbf{GIDS}~\cite{seo2018gids}                  & 97.3               & 99.5        & 98.3   & -    \\ 
& DCNN~\cite{song2020vehicle}                 & 99.95             & 99.65          & 99.80  & -    \\
& MLIDS~\cite{desta2020mlids}                  & 99.9             & 99.9          & 99.9  & -     \\
& HyDL-IDS~\cite{lo2022hybrid}                  & 99.98               & 99.88          & 99.93  &     \\
& \textbf{NovelADS}~\cite{agrawal2022novelads}                  & 99.99               & 100         & 100  & -     \\
& TCAN-IDS~\cite{cheng2022tcan}                  & 99.96             & 99.89          & 99.22  & -     \\
& \textbf{iForest}~\cite{de2021efficient}                  & 95.07                & 99.93          & 97.44  &    -  \\
& GRU~\cite{ma2022gru}                  & 99.32             & 99.13               & 99.22  & -     \\
& \textbf{QCAE-IDS}            &    \textbf{99.63}         &   \textbf{99.38}      &  \textbf{99.50}  &   \textbf{0.25}  \\


\bottomrule
\end{tabular}}
\label{table:comp1}
\end{table}

\begin{table}[t!]
\centering
\caption{Accuracy metric comparison (\%) of our QCAE accelerator against the reported literature on the RPM and Gear spoofing attacks.}
\scalebox{1}{
\begin{tabular}{@{}llllll@{}}
\toprule
\textbf{Attack}  & \textbf{Model} & \textbf{Precision} & \textbf{Recall} & \textbf{F1}  & \textbf{FNR} \\
\midrule
\multirow{5}{*}{RPM} & \textbf{GIDS}~\cite{seo2018gids}                  &    98.3            &    99      & 98.6  &   -   \\ 
& DCNN~\cite{song2020vehicle}                  & 99.99                & 99.94          & 99.96  & -     \\
& MLIDS~\cite{desta2020mlids}                  & 100                & 100          & 100  & -     \\
& HyDL-IDS~\cite{lo2022hybrid}                  & 100               & 100          & 100  & -    \\
& \textbf{NovelADS}~\cite{agrawal2022novelads}                  & 99.9               & 99.9          & 99.9  & -     \\
& TCAN-IDS~\cite{cheng2022tcan}                  & 99.9             & 99.9          & 99.9 & -     \\
& \textbf{iForest}~\cite{de2021efficient}                  & 98.9                &   100       &  99.4 &  -    \\ 
& \textbf{QCAE-IDS}                  &      \textbf{100}        &   \textbf{99.07}            &  \textbf{99.53}  &   \textbf{0} \\
\midrule
\multirow{5}{*}{Gear} & \textbf{GIDS}~\cite{seo2018gids}                  &     98.1            &     96.5     &  97.2  & -    \\ 
& DCNN~\cite{song2020vehicle}                 & 99.99             & 99.89          & 99.94  & -     \\
& MLIDS~\cite{desta2020mlids}                  & 100             & 100          & 100  & -     \\
& HyDL-IDS~\cite{lo2022hybrid}                  & 100               & 100          & 100  & -     \\
& \textbf{NovelADS}~\cite{agrawal2022novelads}                  & 99.99               & 99.8         & 99.9  & -     \\
& TCAN-IDS~\cite{cheng2022tcan}                  & 99.9             & 99.8          & 99.9  & -     \\
& \textbf{iForest}~\cite{de2021efficient}                  & 94.7                & 100          & 97.3  &    -  \\
& \textbf{QCAE-IDS}            &    \textbf{99.91}          &      \textbf{99.40}     &  \textbf{99.66}  &   \textbf{0.12}  \\
\bottomrule
\end{tabular}}
\label{table:comp2}
\end{table}

\subsection{Inference latency}
We quantify the processing latency of the model, starting from the arrival of the CAN message at the interface to determine the detection delay incurred by the approach.
Table~\ref{table:latcomp} compares our result against other approaches in the literature, which utilise different platforms (GPUs, Jetson edge accelerators, Raspberry Pi) and approaches (block of CAN messages v/s individual messages). 
The tightly integrated QCAE-IDS performs inference in 0.43\,ms (B4096 DPU) for a block of 100 CAN frames, which is a 1.3$\times$ improvement over the dedicated line-rate MTH-IDS ECU proposed in~\cite{yang2021mth}.
We also observe the inference latency of the (fp32) model on the Jetson Nano GPU to be 1.8\,ms which is 4.18$\times$ slower than the proposed QCAE-IDS on the hybrid FPGA-based ECU.
The ping-pong buffer at the PL allows the accumulation of 100 CAN IDs (approx 1\,ms at full utilisation of CAN at 1\,Mbps) in one buffer while the previous 100 IDs can be processed from the second buffer by the DPU, effectively hiding the processing latency within the reception delay of a block of messages.
%
The proposed QCAE-IDS hence achieves processing speeds that enable real-time detection of zero-day attack messages on high-speed critical CAN networks.



\begin{table}[t!]
\centering
\caption{Inference latency comparison against other state-of-the-art IDSs (supervised \& unsupervised) reported in literature.}
\scalebox{1}{
\begin{tabular}{@{}lrll@{}}
\toprule
\textbf{Models}     & \textbf{Latency} & \textbf{Frames} & \textbf{Platform} \\ \cmidrule{1-4}
GRU~\cite{ma2022gru} & 890\,ms & 5000 CAN frames & Jetson Xavier NX  \\
MLIDS~\cite{desta2020mlids} & 275\,ms & per CAN frame & GTX Titan X \\
Rec-CNN~\cite{desta2022rec} & 117\,ms & 128 CAN frames & Jetson TX2 \\
\textbf{NovelADS}~\cite{agrawal2022novelads} & 128.7\,ms & 100 CAN frames & Jetson Nano \\
\textbf{GIDS}~\cite{seo2018gids} & 5.89\,ms & 64 CAN frames &  GTX 1080  \\
DCNN~\cite{song2020vehicle} &  5\,ms & 29 CAN frames & Tesla K80 \\
TCAN-IDS~\cite{cheng2022tcan} & 3.4\,ms & 64 CAN frames & Jetson AGX \\
\textbf{MTH-IDS}~\cite{yang2021mth} &  0.574\,ms & per CAN frame & Raspberry Pi 3     \\
\textbf{QCAE-IDS} (ours) & 0.43 ms & 100 CAN frames & Zynq Ultrascale+    \\ 
\bottomrule
\end{tabular}}
\label{table:latcomp}
\end{table}

\subsection{Power consumption and Resource Utilisation}
We further quantify the power consumption and resource overhead (table~\ref{table:dpucomparison}) of our approach to quantify the energy consumption per inference and the hardware resources incurred on the device. 
We find that the B4096 DPU consumes $<$ 30\% LUT \& $\approx$ 40\% DSP resources on the XCZU7EV device, leaving enough resources for other custom accelerators on the PL part of the SoC.
We observe that our model consumed 4.89\,W when measured directly from the device's power rails (using the PYNQ-PMBus package) while performing inference and other tasks on the ECU (with Linux OS), which translates to 2.1\,mJ of energy consumed per inference by the model.
Among other reported results in the literature, our approach results in a $\approx$ 2$\times$ reduction in power consumption when compared to the GRU~\cite{ma2022gru} model, which uses an Nvidia Jetson Xavier as a dedicated IDS node.
We also observed the per-inference energy consumption of our (fp32) CAE model on the Jetson Nano to be 9.36\,mJ when averaged over 10000 inference runs.
Hence, our integrated ECU approach with QCAE-IDS achieves a 4.4$\times$ reduction in energy consumed per inference and a 4.18$\times$ reduction in inference latency over a dedicated Jetson Nano-based IDS accelerator, making our approach a more appropriate choice for deploying IDS in critical CAN systems.

\begin{table}[t!]
\centering
\caption{Performance scaling of the QCAE model using different DPU configurations: impact on per message latency (ms), resource (\%), \& power consumption (W).}
\scalebox{0.85}{
\begin{tabular}{lcccccccc}
\toprule
\multirow{2}{*}{\textbf{DPU}} & \multirow{2}{*}{\textbf{Latency}} & \multicolumn{5}{c}{\textbf{\% Resource Utilisation}} & \multicolumn{2}{c}{\textbf{Power (W)}} \\ \cmidrule{3-9}
 & (ms) & LUT & FF & DSP & BRAM & URAM & Idle & Active \\ \midrule
B512           & 1.1                  &  17.34 & 11.10 &7.18 &16.51 &12.5  & 2.61 & 3.11 \\
B1152           & 0.77                &  19.53&13.92&13.08&19.55&33.33   & 3.12   &  3.56 \\
B4096           & 0.43                &  27.17&25.03&40.74&35.42&47.92   & 4.36   & 4.89  \\
\bottomrule
\end{tabular}}
\label{table:dpucomparison}
\end{table}

%% file: conclusion.tex
\section{Conclusion}\label{sec:conclusion}
In this paper, we present a quantised convolutional autoencoder-based IDS trained only on normal CAN messages capable of detecting zero-day attacks for automotive controller area networks in real time.
The proposed QCAE-IDS achieves state-of-the-art classification accuracy across multiple unseen attack vectors as well as a 1.3$\times$ speed-up in processing latency and $\approx$ 2$\times$ reduction in power consumption with 2.1\,mJ energy per inference when compared to the state-of-the-art IDSs.
The tight integration enables IDS capabilities to be consolidated at the ECU with complete isolation from the ECU function allowing IDS to be efficiently deployed in vehicles. 
In the future, we aim to optimise the proposed IDS to incorporate the payload of the CAN frame as an input feature to achieve higher classification accuracies and make the IDS more robust.
We believe that the proposed IDS architecture can lend itself to low-power real-time IDS architecture for detecting zero-day attacks for emerging vehicular networks like Automotive Ethernet.